\documentclass[a4paper]{article}
\usepackage{amsmath, amsthm}
\usepackage{amssymb}
\usepackage{graphicx}
\usepackage{vpage}
\usepackage[numbers]{natbib}
\setmarginsrb{3.5cm}{4cm}{3.5cm}{4cm}{0pt}{0mm}{0pt}{0mm}

\newtheoremstyle{theorem}
  {10pt}		  
  {10pt}  
  {\sl}  
  {\parindent}     
  {\bf}  
  {. }    
  { }    
  {}     
\theoremstyle{theorem}

\newtheoremstyle{defi}
  {10pt}		  
  {10pt}  
  {\rm}  
  {\parindent}     
  {\bf}  
  {. }    
  { }    
  {}     
\theoremstyle{defi}

\begin{document}

\title{\vspace*{3.5cm} \normalsize \bf STRONGLY NONLINEAR NATURE OF \\INTERFACIAL-SURFACTANT INSTABILITY
  OF COUETTE FLOW}
\author{Alexander L. Frenkel$^1$, David Halpern$^2$ \\
$^1$ University of Alabama, Tuscaloosa, AL 35487-0350, USA \\
email: afrenkel@gp.as.ua.edu \\[2pt]
$^2$ University of Alabama, Tuscaloosa, AL 35487-0350, USA \\
email: dhalpern@as.ua.edu
}
\maketitle
\thispagestyle{empty}

\begin{abstract}
Nonlinear stages of the recently uncovered instability due to insoluble
surfactant at the interface between two fluids are investigated for
the case of a creeping plane Couette flow with one of the fluids a
thin film and the other one a much thicker layer. Numerical simulation
of strongly nonlinear longwave evolution equations which couple the
film thickness and the surfactant concentration reveals that in contrast
to all similar instabilities of surfactant-free flows, no amount of
the interfacial shear rate can lead to a small-amplitude saturation
of the instability. Thus, the flow is stable when the shear is zero,
but with non-zero shear rates, no matter how small or large (while
remaining below an upper limit set by the assumption of creeping flow),
it will reach large deviations from the base values-- of the order
of the latter or larger. It is conjectured that the time this evolution
takes grows to infinity as the interfacial shear approaches zero.
It is verified that the absence of small-amplitude saturation is not
a singularity of the zero surface diffusivity of the interfacial surfactant.


\end{abstract}

\begin{center} \textbf{1. Introduction} \end{center}
Recently, a new instability of a channel flow of two fluid layers
was uncovered in~\citet{FH:02}, \citet{HF:03}. This linear instability was further
investigated in \citet{BP:04}, \citet{BlythPozrikidis:2004}, \citet{Wei:2005a}, \citet{FrenkelHalpern:2005}, \citet{wei2005b}
(\citet{BP:04} and \citet{BlythPozrikidis:2004} included some
nonlinear investigations). It is driven by the interaction of capillary
and Marangoni forces with the interfacial shear of the base flow,
and does not depend on gravity and inertia for its realization. The
instability disappears if the shear is zero. On the other hand, as
was noted in \citet{FH:02}, a nonzero shear is known to enable
the nonlinear saturation of different instabilities, in which capillarity
plays a stabilizing role and the destabilizing factor can be molecular
Van der Waals forces (\citet{BF:83}), or gravity (\citet{BF:83b}, \citet{HF:01}),
or the interfacial jump in viscosity (\citet{HG:85}, \citet{SSBF:85}), or capillary
forces due to the azimuthal curvature of the interface in core-annular
flows (\citet{FB:87}, \citet{papa90}, \citet{HalpernGrotberg:2003}) (see also reviews
\citet{JR:93}, \citet{JosephBaiChenRenardy:1997}). When the shear rate is sufficiently
large, the instabilities saturate with the final amplitude of the
interface undulations being small, so that the whole evolution starting
from small disturbances can be described by a weakly-nonlinear equation
for the film thickness. (The mechanism of this nonlinear saturation
was proposed in \citet{BF:83b} and further clarified in \citet{FB:87}.)
In other cases, the saturated amplitudes for certain regimes are large
and a strongly nonlinear equation is appropriate, as for example for
the flow down a fiber (\citet{FRENKEL:1992}, \citet{FRENKEL:1993},
\citet{KF94}, \citet{KALLIADASISCHANG:1994},
\citet{ChangDemekhin:1999}, \citet{KliakhandlerDavisBankoff:2001}) 
or for the core-annular flow (\citet{FK94}, \citet{K95}, \citet{JosephBaiChenRenardy:1997}).
However, for the same systems, there are other parametric regimes
for which small-amplitude saturation still occurs, so that the weakly-nonlinear
description is good. Since for the instability considered below the
interfacial shear of velocity plays two conflicting roles, working
both for the linear instability and the nonlinear mechanism of saturation
(and also because the dynamics involves a surfactant evolution equation
in addition to the film thickness equation, in contrast to the single-equation
description of evolution in the surfactant-free instabilities), there
is a question whether there are any parametric domains for which the
small-amplitude saturation of the interfacial-surfactant instability
happens (\citet{FH:02}). In this paper, we investigate the nonlinear
stages of the instability with respect to this question (and, to anticipate,
arrive at the conclusion that no small-amplitude saturation occurs,
so that the weakly-nonlinear description inevitably breaks down after
some time).

The rest of the paper is organized as follows. In the next section,
after giving the exact formulation of the Navier-Stokes problem, we
use the lubrication approximation to obtain a simplified system of
coupled equations for the film thickness and the surfactant concentration.
The linear stability governed by these equations is studied in section
3 and the nonlinear evolution is numerically simulated
in section 4. The last section contains the discussion
of results and conclusions.

\begin{center} \textbf{2. Problem formulation} \end{center}

\textbf{2.1 Exact dynamic equations}

\begin{figure}
\begin{center}\includegraphics[%
  width=5in]{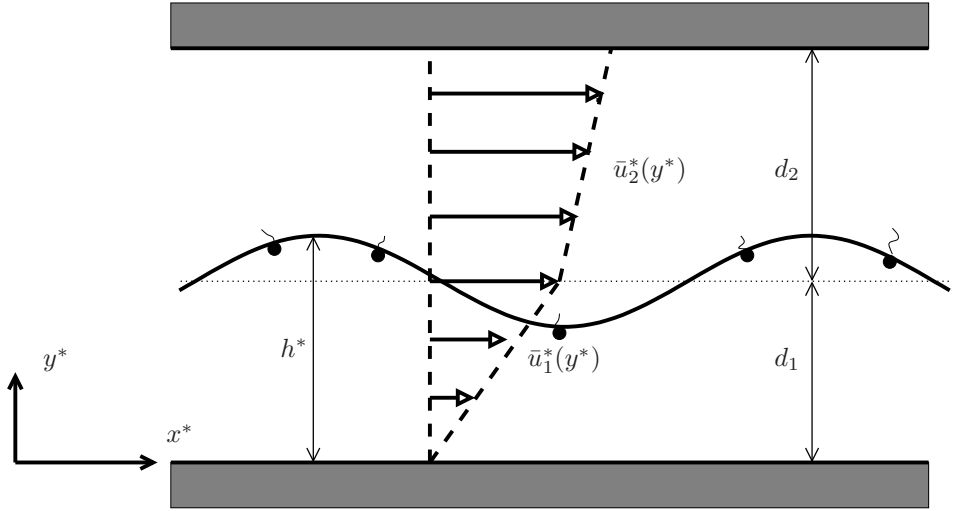}\end{center}
\caption{Definition sketch for a viscous two-layer Couette flow between a
fixed plate at $y^{\ast}=0$ and a moving plate at $y^{\ast}=d_{1}+d_{2}$.
The disturbed interface $y^{\ast}=h^{\ast}(x^{\ast},t^{\ast})$ is
shown as a sinusoidal curve. $\Gamma^{\ast}(x^{\ast},t^{\ast})$ is
the concentration of the insoluble surfactant monolayer shown as discs
with tails. The base velocity profiles are shown by bold arrows.}
\label{fig:fig-geom}
\end{figure}

The formulation of the exact Navier-Stokes problem (as given earlier
in \citet{FH:02}, \citet{HF:03}) is as follows. Consider two immiscible
fluid layers between two infinite parallel plates, as in Fig.~\ref{fig:fig-geom}.
Let the base flow be driven by the combined action of an in-plane
steady motion of one of the plates and a constant pressure gradient
parallel to the plate velocity and directed in either the same or
the opposite sense. It is well known that the base {}``Couette-Poiseuille''
velocity profiles are steady and vary (quadratically) in the spanwise
direction only, and the base interface between the fluids is flat.
 For simplicity, let the densities of the two fluids be equal. Then
gravity does not affect stability of the base flow, and is disregarded
below. Let $y^{\ast}$ be the spanwise, {}``vertical'', coordinate
(the symbol $^{\ast}$ indicates a dimensional quantity). Let the
interface be at $y^{\ast}=d_{1}$ where $d_{1}$ is the thickness
of the thinner layer, and the $y^{\ast}$-axis directed from the thinner
layer ({}``film'') to the thicker one; we will call this the {}``upward''
direction (clearly, since there is no gravity, the notions of {}``up''
and {}``down'' are a matter of convention). Thus, $d_{1}<d_{2}$
holds, where $d_{2}$ is the thickness of the upper fluid. The direction
of the {}``horizontal'' $x^{\ast}$-axis is chosen so that velocity
at the interface is positive, say $U_{1}$ (whereas the velocity at
the lower plate is zero).

The Squire-type theorem for the case with surfactants, proved in \citet{HF:03},
allows us to confine the stability considerations to \textit{two-dimensional}
perturbed flows (in the $x^{\ast}y^{\ast}$-plane). The equation of
the interface is $y^{\ast}=h^{\ast}(x^{\ast},t^{\ast}),$ where $h^{\ast}$
is the film thickness. The Navier-Stokes and incompressibility equations
governing the fluid motion in the two layers are (with $j=1$ for
the lower layer and $j=2$ for the upper one) \[
\rho(\frac{\partial\mathbf{v}_{j}^{\ast}}{\partial t^{\ast}}+\mathbf{v}_{j}^{\ast}\cdot\mathbf{\nabla}^{\ast}\mathbf{v}_{j}^{\ast})=-\mathbf{\nabla}^{\ast}p_{j}^{\ast}+\mu_{j}\mathbf{\nabla}^{\ast2}\mathbf{v}_{j}^{\ast},\quad\mathbf{\nabla}^{\ast}\cdot\mathbf{v}_{j}^{\ast}=0\]
 where $\mathbf{\nabla}^{\ast}=(\frac{\partial}{\partial x^{\ast}},\frac{\partial}{\partial y^{\ast}})$,
$\rho$ is the density (of both fluids), $\mathbf{v}_{j}^{\ast}=(u_{j}^{\ast},v_{j}^{\ast})$
is the fluid velocity with the horizontal component $u_{j}^{\ast}$
and vertical component $v_{j}^{\ast}$, and $p_{j}^{\ast}$ is the
pressure.

We use the {}``no-slip , no-penetration'' boundary conditions (requiring
zero relative velocities) at the plates: $u_{1}^{\ast}=0$, $v_{1}^{\ast}=0$
at $y^{\ast}=0$; and $u_{2}^{\ast}=U_{2}$, $v_{2}^{\ast}=0$ at
$y^{\ast}=d_{1}+d_{2},$ where $U_{2}$ is the velocity of the upper
plate (which can be positive or negative for the general Couette-Poiseuille
flow). The interfacial boundary conditions are as follows. The velocity
must be continuous at the interface: $[\mathbf{v}^{\ast}]_{1}^{2}=0$,
where $[A]_{1}^{2}=A_{2}-A_{1}$ denotes the jump in $A$ across the
interface, i.e. at $y^{\ast}=h^{\ast}(x^{\ast},t^{\ast})$. The interfacial
balances of the tangential and normal stresses taking into account
the jump in the tangential stress (due to the variability of surface
tension) as well as the capillary jump in the normal stress, at $y^{*}=h^{*}(x^{*},t^{*})$,
are \begin{align*}
 & \frac{1}{1+h_{x^{\ast}}^{\ast2}}[(1-h_{x^{\ast}}^{\ast2})\mu(u_{y^{\ast}}^{\ast}+v_{x^{\ast}}^{\ast})+2h_{x^{\ast}}^{\ast2}\mu(v_{y^{\ast}}^{\ast}-u_{x^{\ast}}^{\ast})]_{1}^{2}=-\frac{\sigma_{x^{\ast}}^{\ast}}{(1+h_{x^{\ast}}^{\ast2})^{1/2}}\\
 & [(1+h_{x^{\ast}}^{\ast2})p^{\ast}-2\mu(h_{x^{\ast}}^{\ast2}u_{x^{\ast}}^{\ast}-h_{x^{\ast}}^{\ast}(u_{y^{\ast}}^{\ast}+v_{x^{\ast}}^{\ast})+v_{y^{\ast}}^{\ast})]_{1}^{2}=\frac{h_{x^{\ast}x^{\ast}}^{\ast}}{(1+h_{x^{\ast}}^{\ast2})^{1/2}}\sigma^{\ast}\end{align*}
 where $\sigma^{\ast}$ is the surface tension (the subscripts $x^{\ast},y^{\ast},t^{\ast}$
denote the corresponding differentiation). The kinematic boundary
condition (the conservation of mass condition) is\begin{equation}
\frac{\partial h^{\ast}}{\partial t^{\ast}}+\frac{\partial}{\partial x^{\ast}}\int\limits _{0}^{h^{\ast}(x^{\ast},t^{\ast})}u^{\ast}\, dy^{\ast}=0.\label{eq:kbc}\end{equation}
 The surface concentration of the insoluble surfactant on the interface,
$\Gamma^{\ast}(x^{\ast},t^{\ast})$, obeys the following equation
(see a simple derivation in \citet{HF:03}): \begin{equation}
\frac{\partial(H\Gamma^{\ast})}{\partial t^{\ast}}+\frac{\partial}{\partial x^{\ast}}\left(H\Gamma^{\ast}u^{\ast}\right)=D_{s}\frac{\partial}{\partial x^{\ast}}\left(\frac{1}{H}\frac{\partial\Gamma^{\ast}}{\partial x^{\ast}}\right)\label{eq:gammas}\end{equation}
 where $H=\sqrt{1+h_{x}^{\ast2}}$, and $D_{s}$ is the surface molecular
diffusivity of surfactant; $D_{s}$ is usually negligible, and is
discarded below. We assume that the surfactant concentration $\Gamma^{\ast}(x^{\ast},t^{\ast})$
is sufficiently small, remaining always far below its saturation value,
so that the linear dependence of the surface tension on the surfactant
concentration is a good approximation (see \citet{EdwardsBrennerWasan:1991},
p.~171): $\sigma^{\ast}=\sigma_{0}-E\left(\Gamma^{\ast}-\Gamma_{0}\right)$,
where $\Gamma_{0}$ is the base surfactant concentration$,\sigma_{0}$
is the base surface tension and $E$ is a constant.

We use the following nondimensionalization: \begin{equation}
(x,y)=\frac{(x^{\ast},y^{\ast})}{d_{1}}\;,\; t=\frac{t^{\ast}}{d_{1}\mu_{1}/\sigma_{0}}\;,\;(u,v)=\frac{(u^{\ast},v^{\ast})}{\sigma_{0}/\mu_{1}}\;,\; p=\frac{p^{\ast}}{\sigma_{0}/d_{1}}\;,\;\Gamma=\frac{\Gamma^{\ast}}{\Gamma_{0}}\;,\;\sigma=\frac{\sigma^{\ast}}{\sigma_{0}}.\label{eq:nondim}\end{equation}

In the base flow, the interface is flat, $\overline{h}=1$, and the
surfactant concentration is uniform, $\overline{\Gamma}=1$, where
the overbar indicates a base-state quantity. \ Since the base flow
has a constant pressure gradient, say $2q$, the two layers have equal
pressures, $\overline{p}_{1}=\overline{p}_{2}=2qx$, and the velocity
profiles are\begin{align}
\bar{u}_{1}(y) & =ry+qy^{2},\; & \bar{v}_{1} & =0\;,\; &  & \text{for
$0\leq y\leq1$},\label{eq:vel1}\\
\bar{u}_{2}(y) & =(r+q)\frac{m-1}{m}+\frac{r}{m}y+\frac{q}{m}y^{2},\; & \bar{v}_{2} & =0\;,\; &  & \text{for $1\leq y\leq n+1$},\label{eq:vel2}\end{align}
 where $n=d_{2}/d_{1}$ is the aspect ratio, $n\geq1$, and $m=\mu_{2}/\mu_{1}$.
\ The pair of constants $r$ and $q$ is used in place of the physical
control parameters, the relative velocity of the plates and the pressure
gradient, to characterize the base flow. \ The base interfacial velocity
is $w=r+q,$ so $r+q>0.$ The parameter $s=r+2q$ is the interfacial
shear of velocity (on the lower-fluid side).

\textbf{2.2 Longwave (lubrication) approximation}

We will assume that the aspect ratio is large, $n>>1$ (actually we
work in the zeroth order of the small parameter $1/n$, so the parameter
$n$ disappears from the equations). For sufficiently slow flow, the
inertia terms in the Navier-Stokes equations are negligible. In the
well-known lubrication approximation, which implies that the streamwise
characteristic lengthscale is much larger than the film thickness
(see, e.g., the review papers \citet{FI96} and \citet{ODB:97}),
the simplified dynamic equations are, in the dimensionless form,

\begin{align}
\frac{\partial p}{\partial x} & =\frac{\partial^{2}u}{\partial y^{2}},\label{eq:xmom}\\
\frac{\partial p}{\partial y} & =0,\label{eq:ymom}\\
\frac{\partial v}{\partial y} & =-\frac{\partial u}{\partial x},\label{eq:continuity}\end{align}
 where $u=u_{1}-\bar{u}_{1}$, $v=v_{1}-\overline{v}_{1},$ and $p=p_{1}-\overline{p}_{1}$.

The no-slip no-penetration boundary conditions at the lower plate
prescribe that\[
u=0=v\text{ \,\, at \,\,}y=0.\]
 As usual for this class of systems with infinitely large aspect ratio
(e.g., \citet{BF:83}, \citet{BF:83b}, \citet{H:85}, \citet{FB:87}), the disturbance quantities
of the thick layer are negligible in the interfacial stress boundary
conditions (at least if the viscosity ratio is not too different from
one, so that at $y=h(x,t)$ the difference of base velocities is not
much larger than $u$, and hence $u$ and the corresponding velocity
disturbance in the thick layer are of the same order of magnitude),
so the simplified boundary conditions at $y=h(x,t)$ for the system
(\ref{eq:xmom}-\ref{eq:continuity})\ are\[
p=-[1-M(\Gamma-1)]h_{xx}\]
 (the curvature term is retained to allow for the possibility of large
surface values of tension) and \[
u_{y}=-M\Gamma_{x}\]
 where the Marangoni number $M$ is defined as $M=E\Gamma_{0}/\sigma_{0}$,
and we have neglected $h_{x}^{2}$ (as compared to 1) as well as some
other terms, in accordance with the longwave (small-slope) approximation.
\ The solution for $u$ in terms of $h(x,t)$ and $\Gamma(x,t)$
is \[
u=\{ h[(1+M-M\Gamma)h_{xx}]_{x}-M\Gamma_{x}\} y-\frac{1}{2}[(1+M-M\Gamma)h_{xx}]_{x}y^{2}.\]
 We substitute the total velocity $u_{1}=ry+qy^{2}+u$ into (i) the
dimensionless mass conservation equation (\ref{eq:kbc}), $h_{t}+[\int\limits _{0}^{h(x,t)}u_{1}(x,y,t)\, dy]_{x}=0,$
and (ii) the simplified surfactant conservation equation\[
\Gamma_{t}+[\Gamma(rh+qh^{2}+u)]_{x}=0\]
 (with $u$ being evaluated at the interface $y=h(x,t)$), where we
have set $H=1$ \ (since $h_{x}^{2}<<1$) and the right-hand side
of equation (\ref{eq:gammas}) to zero since the molecular diffusivity
$D_{s}$ is usually very small. As a result, we obtain a system of
two coupled evolution equations, for $h$ and $\Gamma$: \begin{align}
h_{t}+\left(r\frac{h^{2}}{2}+q\frac{h^{3}}{3}-M\Gamma_{x}\frac{h^{2}}{2}+\frac{1}{3}[(1+M-M\Gamma)h_{xx}]_{x}h^{3}\right)_{x} & =0,\label{eq:hevol}\\
\Gamma_{t}+\left[\Gamma\left(rh+qh^{2}-M\Gamma_{x}h+\frac{1}{2}h^{2}[(1+M-M\Gamma)h_{xx}]_{x}\right)\right]_{x} & =0.\label{eq:gammaevol}\end{align}
 The terms containing $r$ and $q$\ are clearly due to the nonzero
base velocity.

To simplify, we make some additional assumptions: We assume $q=0$
(then the velocity is piecewise-linear and $r$ is its slope in the
film; also $r$ is equal to the interfacial velocity) and $M<<1$,
and also $M\Gamma<<1$. Rescaling $\tilde{x}=\beta x$ and $\tilde{t}=r\beta t$
where $\beta=(3M/2)^{1/2}$, introducing the constant $C=\beta M/2r=(3M^{3}/(8r^{2}))^{1/2}$,
and dropping the tildes from the new variables, the equations take
the form \begin{align}
h_{t}+[\frac{h^{2}}{2}+C(-\Gamma_{x}h^{2}+h_{xxx}h^{3})]_{x} & =0,\label{eq:evolh}\\
\Gamma_{t}+\left[\Gamma\left(h+C(-2\Gamma_{x}h+\frac{3}{2}h^{2}h_{xxx})\right)\right]_{x} & =0.\label{eq:evolg}\end{align}
 In each of these equations, the second term is due to the interfacial
shear, the third one is due to Marangoni forces, and the last one
describes the effect of capillarity.

In the rescaled variables, the longwave requirement reads \[
|\beta h_{x}|<<1\]
 which, assuming $M^{1/2}<<1$, can be satisfied even in some cases
with $h_{x}\gtrsim1$ (note that the unit of length for $x$ is much
greater than that for $h$).

Note that because of the conservative form of the equations (\ref{eq:evolh})
and (\ref{eq:evolg}), the integral quantities $\int h(x,t)dx$ and
$\int\Gamma(x,t)dx$ are time-independent if they are taken over domains
with periodic boundary conditions.

We note that it is not difficult to obtain weakly-nonlinear evolution
equations for the disturbances $\eta$ and $g$ defined by \[
h=1+\eta,\,\,\,\Gamma=1+g.\]
 One assumes the disturbances to be finite but small,\[
\eta<<1,\;\, g<<1,\]
 and retains only the leading-order linear and nonlinear terms in
$\eta$ and $g$ in equations (\ref{eq:evolh}) and (\ref{eq:evolg}).
Changing to the coordinate $z=x-t$ (which eliminates the term $\eta_{x}$
from equation (\ref{eq:evolh}) and the term $g_{x}$ from equation
(\ref{eq:evolg})), we obtain\begin{align}
\eta_{t}+\left(\frac{\eta^{2}}{2}-Cg_{z}+C\eta_{zzz}\right)_{z} & =0,\label{eq:res-w-n-h}\\
g_{t}+\left(\eta g+\eta-2Cg_{z}+\frac{3C}{2}\eta_{zzz}\right)_{z} &
=0.\label{eq:res-w-n-g}\end{align} 
 (In this derivation, for example the term proportional to $g_{z}\eta_{z}$
coming from the third term in Eq.~(\ref{eq:evolh}) is smaller by
a factor of $\eta<<1$ than the term proportional to $g_{zz}$ arising
from the same term in Eq.~(\ref{eq:evolh}). Therefore, the smaller
term has been dropped, while, on the other hand, the term with
$({\eta^{2}})_{z}$ 
has been retained in Eq.~(\ref{eq:res-w-n-h}) as the leading order
nonlinear term since there is no other term which would be clearly
larger than it: Indeed, depending on the given conditions, the parameter
$C$ can take values ranging from very small to very large, and the
same is true of the characteristic length scale measured in the units
we have chosen. The only restriction on $C$ is imposed by the requirement
of smallness for the modified Reynolds number, $Red_{1}/\Lambda<<1$,
where $Re$ is the Reynolds number based on the interfacial velocity
and the thickness $d_{1}$, and $\Lambda/d_{1}$ is the dimensionless
streamwise length scale of solutions. This gives the constraint
$C>>M^{3/2}\rho\sigma_{0}d_{1}^{2}/(\mu_{1}^{2}\Lambda)$. The
nonlinear term in Eq.~(\ref{eq:res-w-n-g}) is clearly smaller than
the linear term $\eta_z$ in view of $g<<1$; it could be omitted and
has only been retained in order to show the leading nonlinear term.)

These weakly-nonlinear evolution equations can be rescaled to a {}``canonical
form'' having just one parametric coefficient: \begin{align}
\eta_{t}+\eta\eta_{z}-g_{zz}+\eta_{zzzz} & =0,\label{eq:etaeq}\\
g_{t}-g_{zz}+\frac{3}{4}\eta_{zzzz}+A\eta_{z} & =0\label{eq:geq}\end{align}
 (where the parameter $A$ is inversely proportional to the parameter
$C$). {[}We notice that these weakly-nonlinear equations have certain
similarities with the model system constructed for specific purposes
in \citet{MalomedFengKawahara:2001}: Both systems are first
order in time and fourth order in space and have similar nonlinear
terms, although couplings are different. For a review of other systems
of coupled evolution equations, see, e.g., \citet{GlasmanGolovinNepomnyashchy:2004}.{]}
However, the above weakly-nonlinear equations turn out to be good
for just a short transient stage: as we will see below, the small-amplitude
saturation does not take place and disturbances quickly grow beyond
the scope of the weakly-nonlinear equations. To repeat, this absence
of the small-amplitude saturation in the interfacial-surfactant instability,
contrasting with the presence of such small-amplitude saturation in
all previously known surfactant-free instabilities of two-fluid flows,
is the main point of the present communication.

\begin{center}
\textbf{3. Normal modes of infinitesimal disturbances}
\end{center}

\begin{figure}
\begin{center}\includegraphics[%
  height=3in]{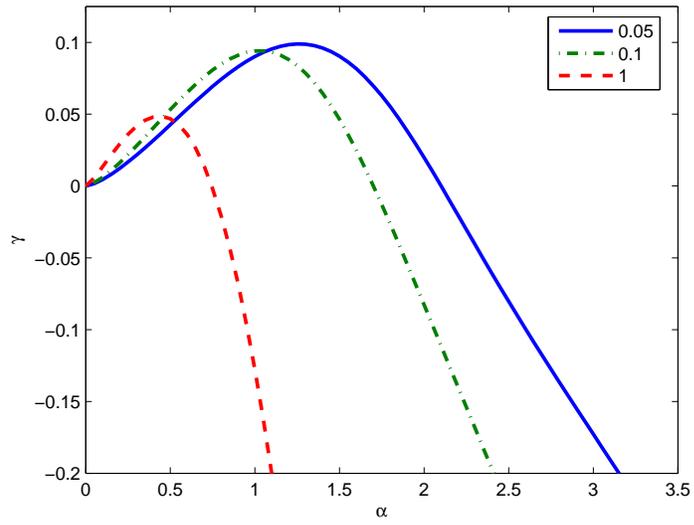}\end{center}
\caption{Dependence of the growth rate $\gamma$ on the wavenumber $\alpha$
of a normal mode for three values of the parameter $C$ indicated
in the legend.}
\label{fig:fig-disp1}
\end{figure}

\begin{figure}
\begin{center}\includegraphics[%
  height=3in]{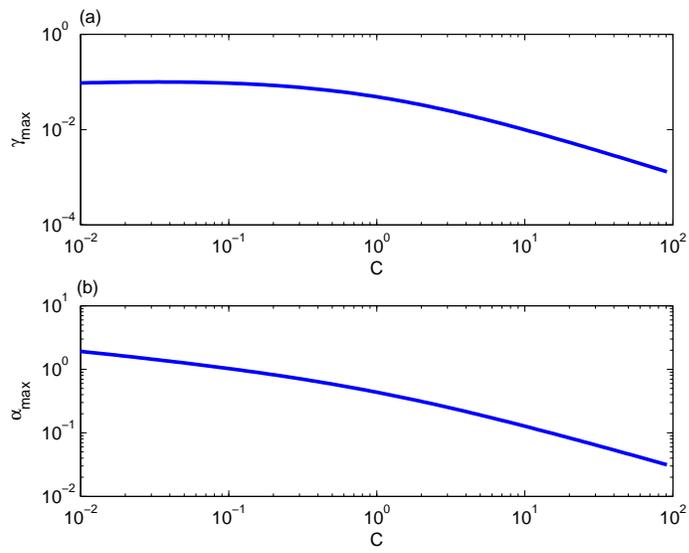}\end{center}
\caption{(a) Growth rate and (b) wavenumber of the fastest-growing normal
mode versus $C$.}
\label{fig:fig-disp2}
\end{figure}

To examine behavior of infinitesimal disturbances to the steady uniform
flow with $\Gamma=1$ and $h=1$, we substitute $h=1+\eta$ and $\Gamma=1+g$
into Eqs.~(\ref{eq:evolh})-(\ref{eq:evolg}) , and neglect the nonlinear
terms, obtaining the linear system \begin{align}
\eta_{t}+\eta_{x}-Cg_{xx}+C\eta_{xxxx} & =0,\label{eq:etaleq}\\
g_{t}+g_{x}+\eta_{x}-2Cg_{xx}+\frac{3}{2}C\eta_{xxxx} & =0.\label{eq:gleq}\end{align}
 It is known (\citet{FH:02}, \citet{HF:03}) that the normal modes $(\eta,g)\propto e^{i\alpha x}e^{\gamma t}e^{i\omega t}$
where $\alpha$ is the wavenumber, $\gamma(\alpha)$ is the growth
rate, and $\omega(\alpha)$ is the time frequency, are unstable--if
and only if $C<\infty$ (i.e., the shear velocity $r\ne0$)--for {}``long''
waves, $0<\alpha<\alpha_{0}$, where $\alpha_{0}$ is a marginal wavenumber.
The longwave asymptotic dependence of the growth rate is $\gamma\propto\alpha^{3/2}$
as $\alpha\rightarrow0$. The normal modes decay if $\alpha>\alpha_{0}$,
with $\gamma\sim-\frac{1}{4}\alpha^{2}$ as $\alpha\rightarrow\infty$.
(We note that the decay is weaker than $\alpha^{4}$ despite the presence
of the fourth-derivative terms in the linearized equations (\ref{eq:etaleq})
and (\ref{eq:gleq})). Typical dispersion curves, the growth rate
versus wavenumber, are shown in Fig.~\ref{fig:fig-disp1}. They have
the maximum value $\gamma_{max}$ at a certain wavenumber $\alpha_{max}$
depending on the value of the equation parameter $C$. These dependences
are shown in Fig.~\ref{fig:fig-disp2}.

Multiplying Eq.~(\ref{eq:etaleq}) by $\eta$ and (\ref{eq:gleq})
and integrating over the interval of spatial periodicity, and taking
into account that $\eta$ and $g$ are out of phase by nearly $\pi$
in unstable modes, shows that the fourth derivative (capillary) term
is stabilizing in Eq.~(\ref{eq:etaleq}) but destabilizing in Eq.~(\ref{eq:gleq}).
Similarly, the Marangoni term, the one with $g_{xx}$, is stabilizing
for $g$ but destabilizing with regard to $\eta$. %
\newpage
\begin{center}
\textbf{4. Nonlinear numerical simulations}
\end{center}

We solved the system of equations (\ref{eq:evolh})
and (\ref{eq:evolg}) numerically on $0\leq x\leq2\pi Q$ with periodic
boundary conditions using the method of lines. This involved the approximation
of the spatial derivatives with finite differences and an implicit
Gear's method for the time-stepping. The number of nodes in the spatial
discretization ranged from hundreds to tens of thousands. Also, some
results were checked using standard pseudospectral Fourier method
(see e.g. \citet{KF94}) in conjunction with an implicit Gear's
method for the time-stepping. The number of modes ranged from 256
to 1024 for different cases. We have used dealiasing as described
e.g. in \citet{KF94} (similar to \citet{CanutoHussainiQuarteroniZang:1987}),
filtering out slightly more than three-fifths of the modes with higher
wavenumbers from the unknown functions and their derivatives, computed
in the Fourier space, before computing the nonlinear terms of the
equations in the physical space. In all cases, we took care that the
length of the interval determined by the parameter $Q$ was sufficiently
large to accommodate five or more elementary pulses, as it is known,
e.g. for the Kuramoto-Sivashinsky equation, that the results can be
sensitive to the type of boundary conditions and the length of the
interval when the latter is not sufficiently large (\citet{PS:91},
\citet{FI96}, \citet{WH:99}). 
{[}In particular, the preliminary nonlinear saturation results found--although
for very different parametric regimes--in \citet{BP:04}, may
not hold for sufficiently large domains. We have checked that such
was indeed the case for our semi-infinite flow configuration: The
small-amplitude saturation occurring for a narrow band of values of
$Q$ near criticality, such that the periodicity 
interval accommodated just one unstable normal mode, disappeared
for larger values of $Q$, since the corresponding larger intervals
can accommodate additional longwave modes which are unstable. Of course,
to any one-pulse solution for a small $Q=a$ there is a corresponding
solution for an $n$ times larger value, $Q=na$, consisting of $n$
identical $a$-pulses. However, as a rule, this $n$-pulse solution
is unstable to longwave disturbances and therefore is not observed
as an outcome of evolution. Conversely, the results are essentially
insensitive to further increase of $Q$ chosen in this way.{]}

\begin{figure}
\begin{center}\includegraphics[%
  width=6in]{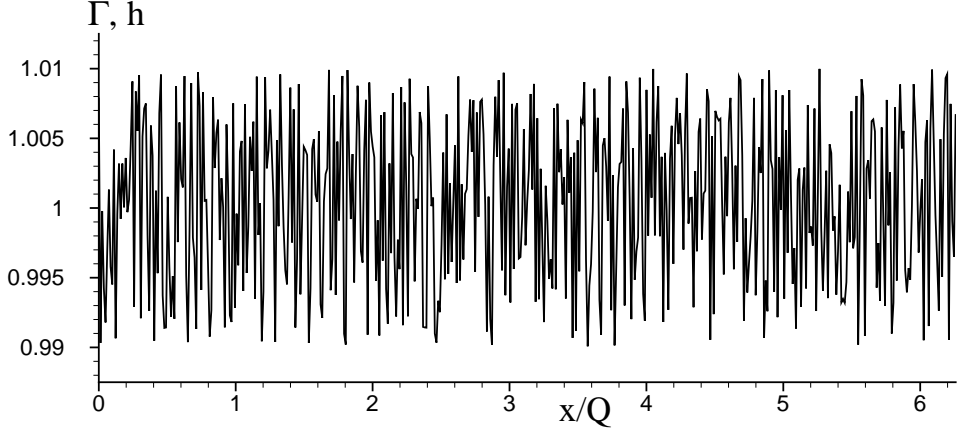}\end{center}

\caption{Typical initial profile of surfactant concentration or film thickness:
a white noise distribution, with deviation amplitude of $10^{-2}$
for $512$ modes.}

\label{fig:fig-initbc}
\end{figure}

For the initial conditions we use small-amplitude white-noise deviations
from the base uniform profiles. The deviation values at the spatial
nodes are chosen from the uniform random number distribution on the
interval $[-10^{-2},10^{-2}]$. Typical initial profiles are shown
in Fig.~\ref{fig:fig-initbc}.

\begin{figure}
\includegraphics[%
  width=2.5in,
  keepaspectratio]{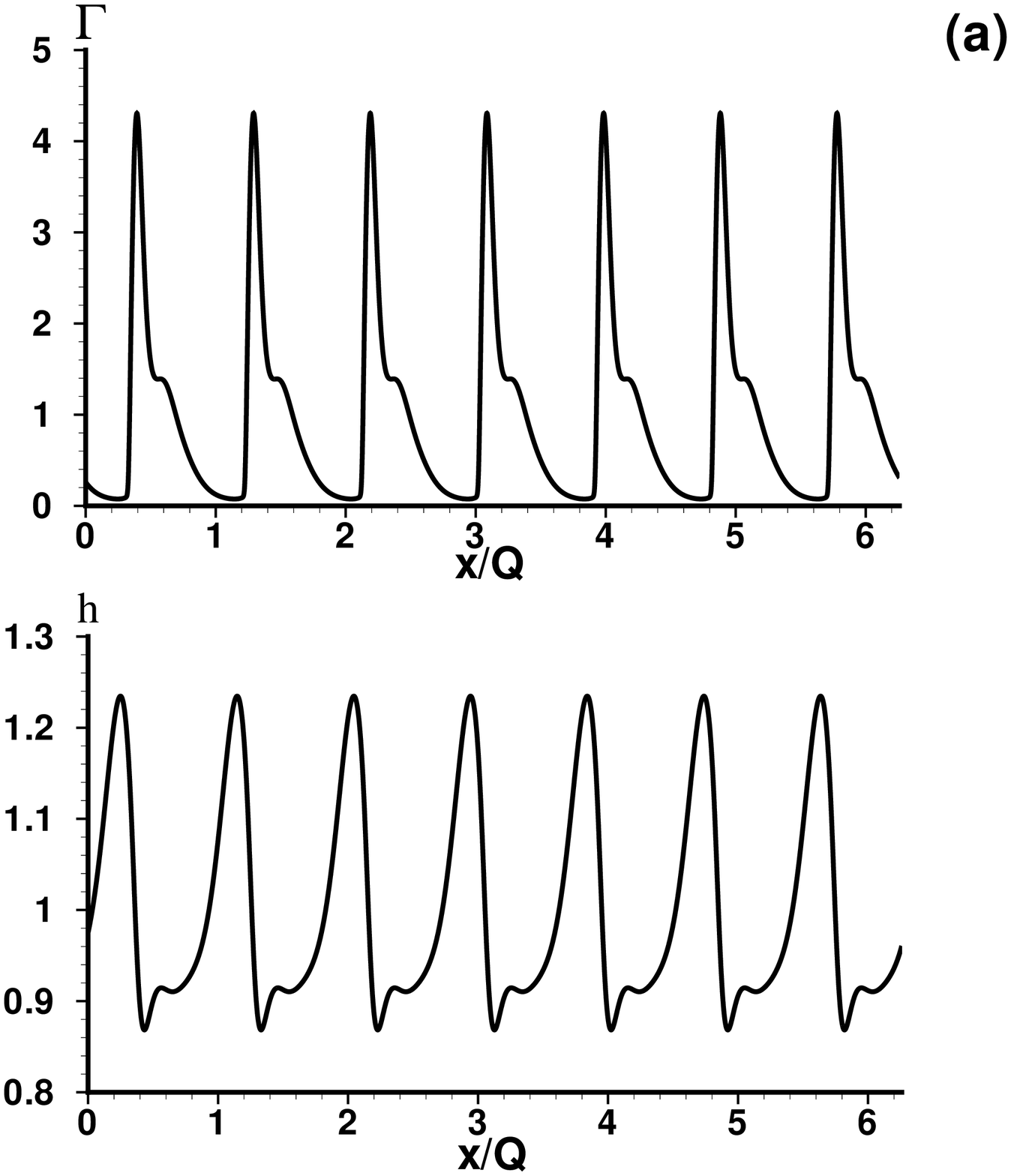}
\includegraphics[%
  width=2.5in,
  keepaspectratio]{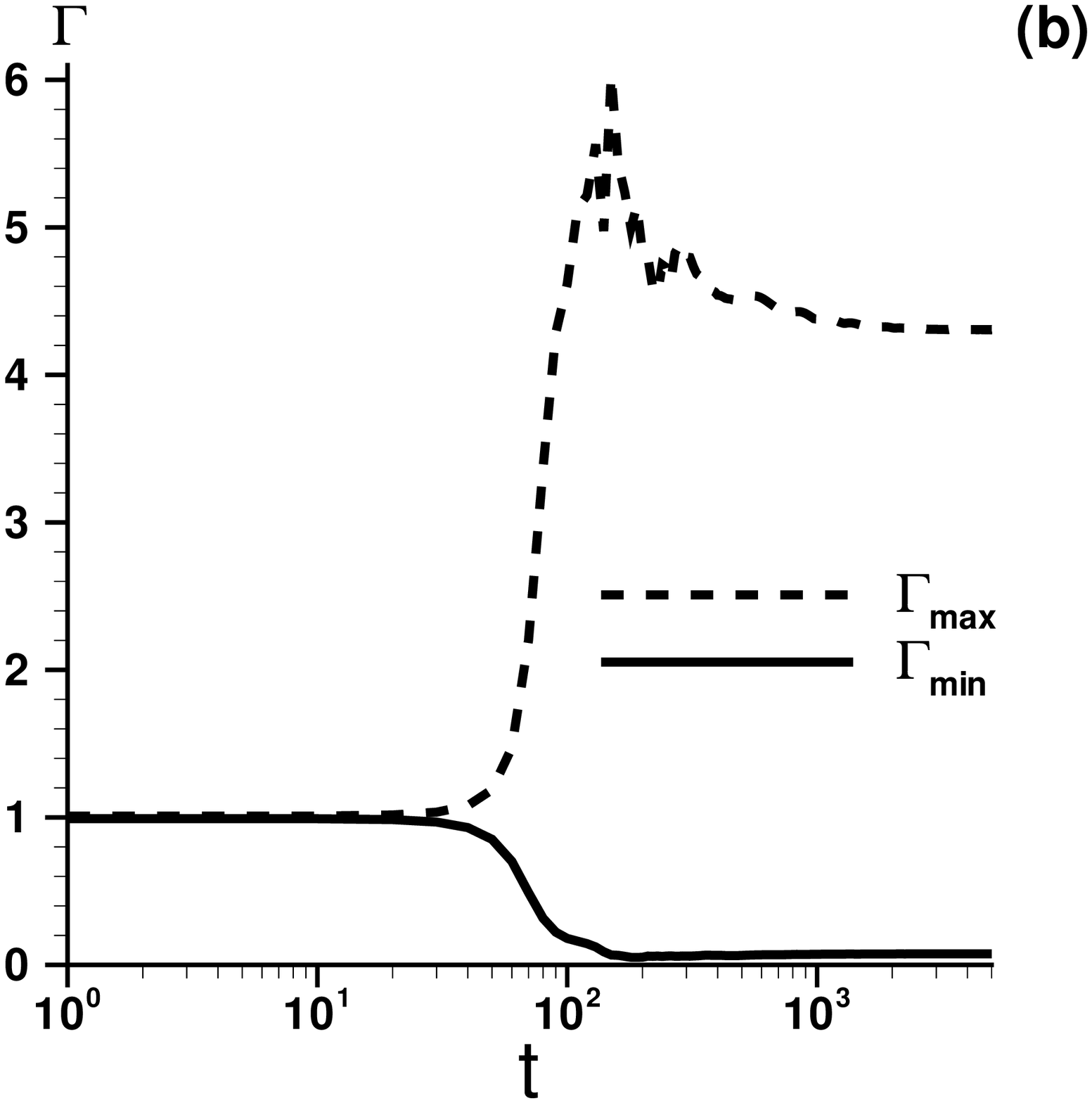}
\hspace*{7.5cm}
\includegraphics[%
  width=2.5in,
  keepaspectratio]{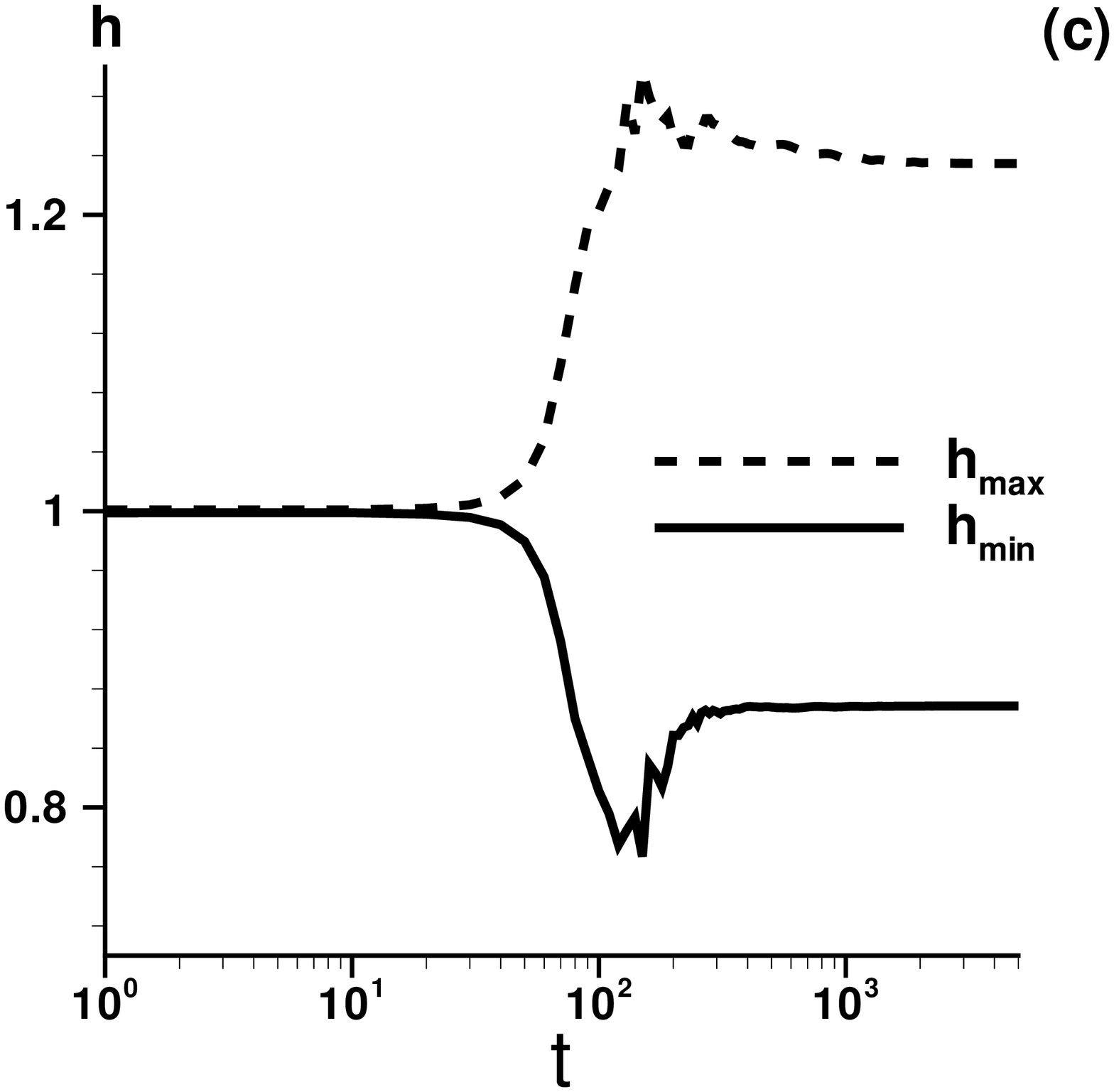}

\caption{\label{fig:saturatedsol-cp01q5p2}A typical evolution of the surfactant
concentration and film thickness. (a) The saturated travelling-wave
profiles of the surfactant concentration and the film thickness, at
$t=5000$. (b) The time dependence of the minimum and maximum (over
$x$) surfactant concentration. (c) The time dependence of the minimum
and maximum film thickness. Here $C=10^{-2}$ (and $Q=5.2$).}
\end{figure}

\begin{figure}
\begin{center}\includegraphics[%
  width=2.5in,
  keepaspectratio]{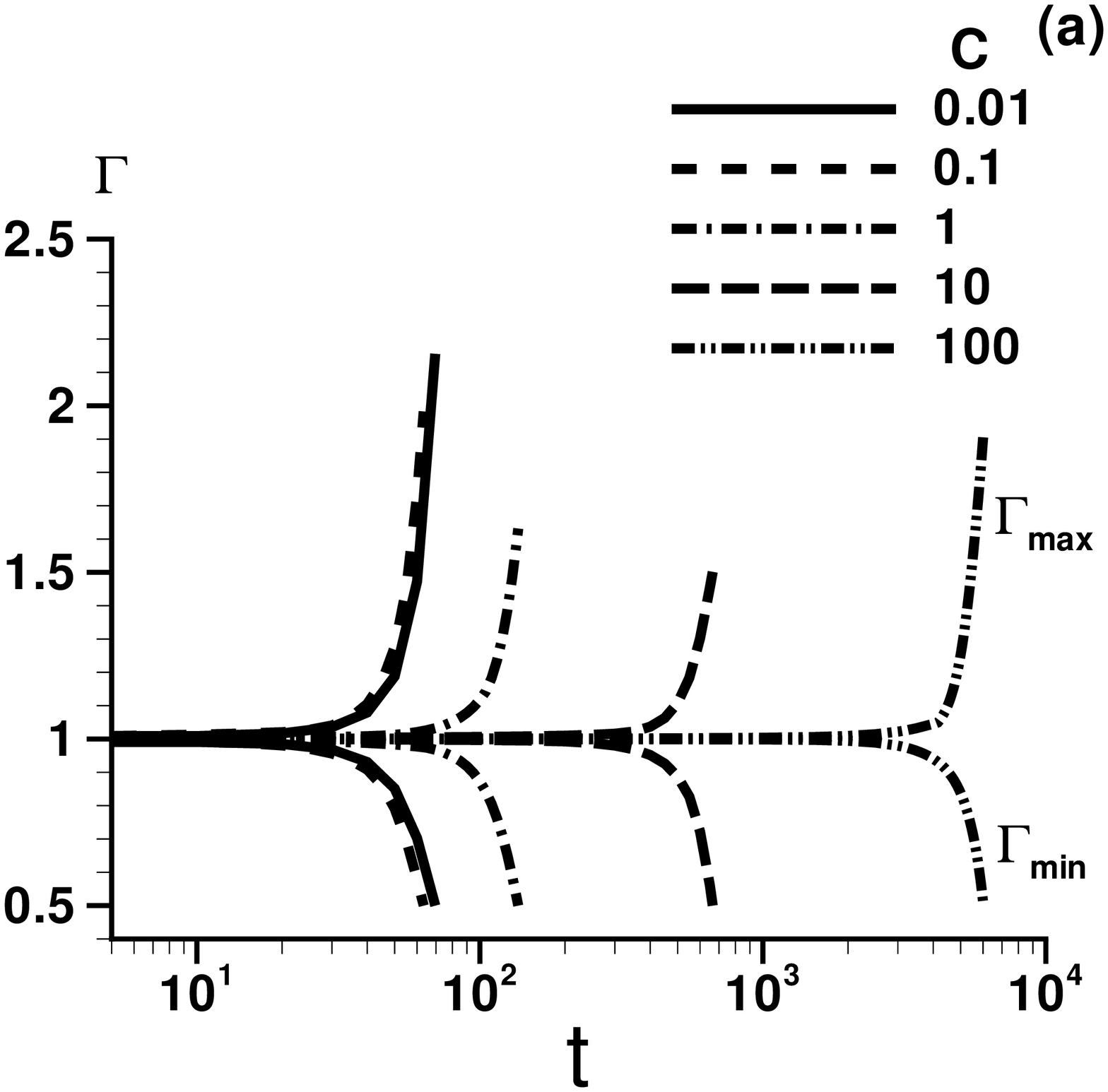}\includegraphics[%
  width=2.5in,
  keepaspectratio]{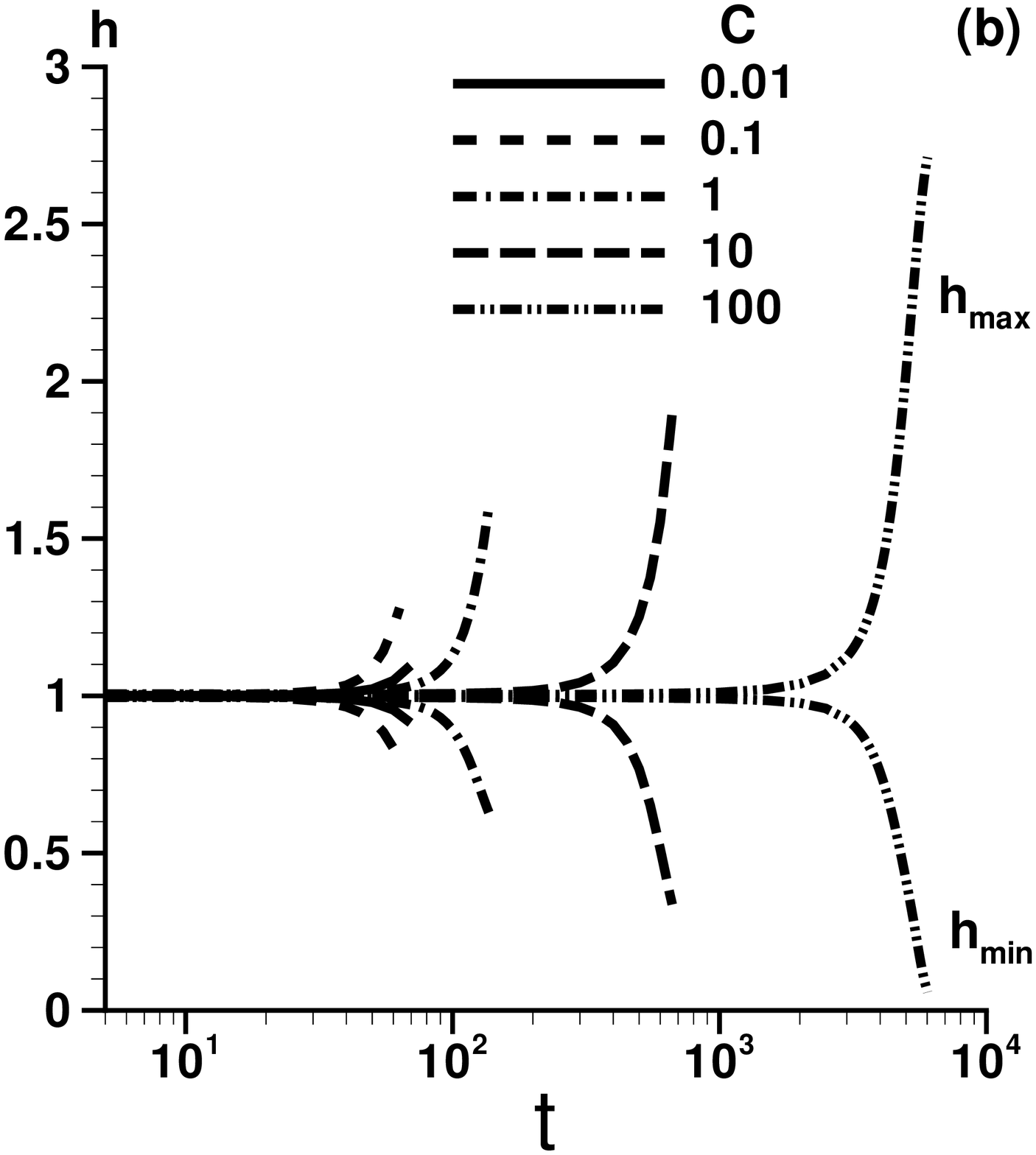}\end{center}

\caption{Evolution of (a) surfactant concentration and (b) film thickness:
The dependence of the minimum and maximum (over $x$) values $\Gamma_{max}$,
$\Gamma_{min}$ and $h_{max}$, $h_{min}$ on time $t$, for the five
different values of $C$ shown in the legend. Each evolution is shown
up to the time at which the minimum surfactant concentration reaches
the value of $0.5$. (In the order of increasing $C$, the values
of $Q$ used in these simulations are $5.2$, $8$, $18$, $50$,
$250$.)}

\label{fig:fig-evolminmaxgh-manyc}
\end{figure}

In the subsequent evolution, the shortwave normal modes die out, but
the longwave ones grow, and the spatial profiles evolve to the pulse
trains, like the travelling waves shown in Fig. \ref{fig:saturatedsol-cp01q5p2}(a).
A typical evolution of the instantaneous global maxima and minima
of the spatial profiles of the surfactant concentration $\Gamma$
and the film thickness $h$ is shown in Fig. \ref{fig:saturatedsol-cp01q5p2}(bc):
There, the deviations of $\Gamma$ and $h$ saturate but at levels
which are not small as compared to non-disturbed values. In Fig. \ref{fig:fig-evolminmaxgh-manyc},
the evolution for different parameter values is shown to emphasize
the point that the deviations in all cases grow to levels of order
one. To this end, each time series is followed up to the instant at
which the minimum surfactant concentration reaches the representative
level of 0.5. We used five different values of the equation parameter
$C$, from small to order one to large: $10^{-2}$, $10^{-1}$, $1$,
$10$, and $10^{2}$, with sufficiently large values of $Q$ (as discussed
above; also, in all cases, we have checked that refining the spatial
and temporal resolutions does not change the results.). The figure
clearly shows that in all cases the deviations of the surfactant concentration
and the film thickness grow and become non-small in comparison with
their base values. As a result, the weakly nonlinear equations (\ref{eq:etaeq})
and (\ref{eq:geq}), based on the assumption that the deviations are
small, would cease to be good. (Also, clearly, if the film thickness
becomes sufficiently small as for $C=100$ (see Fig. \ref{fig:fig-evolminmaxgh-manyc}(b)),
the molecular van der Waals forces should be taken into account.)
The greater the value of $C$, the larger time (of the governing equations)
the evolution takes. 


\begin{figure}
\begin{center}\includegraphics[%
  width=5in,
  keepaspectratio]{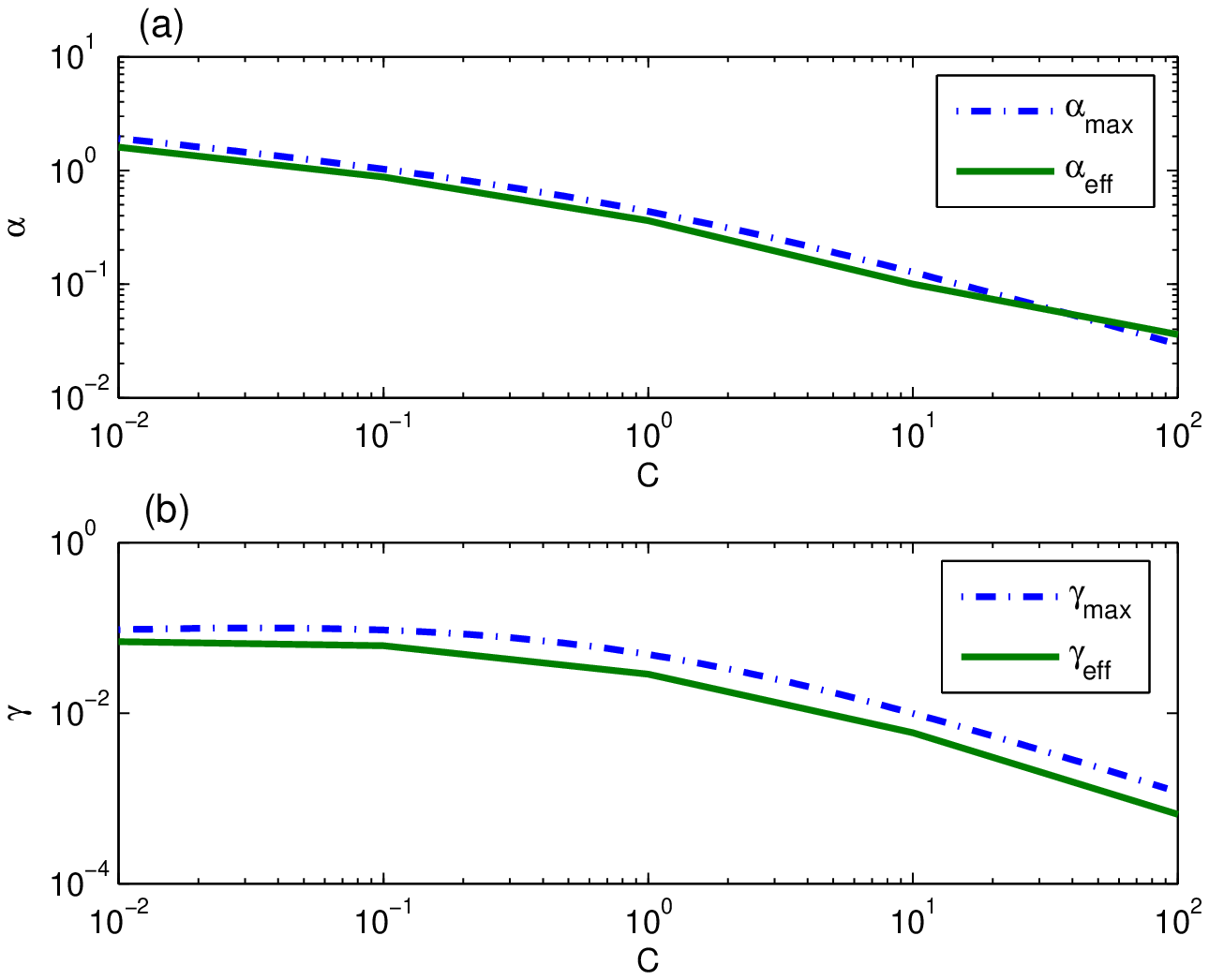}\end{center}

\caption{$C$-dependence of (a) the effective wavenumber and (b) the effective
growth rate (defined in the text). For comparison the corresponding
linear-theory quantities $\gamma_{max}$ and $\alpha_{max}$ are included.}

\label{fig:effective-ga}
\end{figure}
 As was mentioned above, the initial configuration with many maxima
and minima (see Fig.~\ref{fig:fig-initbc}) quickly evolves, by coalescences,
to just a few pulses (see Fig.~\ref{fig:saturatedsol-cp01q5p2}a).
Another way to describe this is to note that the normal modes with
large wavenumbers are damped and die out. The normal mode of the fastest
growth has a tendency to dominate, at least initially. We can define
an effective wavenumber $\alpha_{eff}$ of an instantaneous profile
to be $2\pi$ over the {}``length per one pulse''. The latter is
the interval length $2\pi Q$ divided by the number of pulses $N$.
Therefore, $\alpha_{eff}=N/Q$. The dependence of $\alpha_{eff}$
on $C$ is shown in Fig.~\ref{fig:effective-ga}(a) together with
that of the wavenumber $\alpha_{max}$ of the fastest growing normal
mode. These are seen to be close throughout the entire range of $C$
shown in Fig.~\ref{fig:effective-ga}(a).

We also define an effective growth rate $\gamma_{eff}$ to be that
of a (hypothetical) normal mode that would have the same minimum surfactant
concentration as the actual nonlinear profile at two instances of
time, $t=0$ and $t=t_{1}$, where $t_{1}$ is determined by $\Gamma_{min}(t_{1})=0.5$.
This gives $\gamma_{eff}=\log(0.5/10^{-2})/t_{1}$, that is $\gamma_{eff}=\log(50)/t_{1}$.
Fig.~\ref{fig:effective-ga}(b) shows $\gamma_{eff}$ (inversely
proportional to the evolution time $t_{1}$) as a function of $C$.
For comparison, the growth rate $\gamma_{max}$ of the fastest growing
normal mode is included in the figure. $\gamma_{eff}$ is always smaller
than $\gamma_{max}$ because the initial exponential growth of small
disturbances slows down as the magnitude of the disturbances grows.
The figure also reflects on the fact that the evolution time grows
with $C$. We conjecture that $t_{1}\rightarrow\infty$ as $C\rightarrow\infty$--
that is, (assuming that $M$ is fixed), as the shear rate approaches
zero. (It is easy to see that this would mean that the corresponding
physical, dimensional, time of the evolution diverges to infinity
as well.)

\begin{figure}
\begin{center}\includegraphics[%
  width=2.5in,
  keepaspectratio]{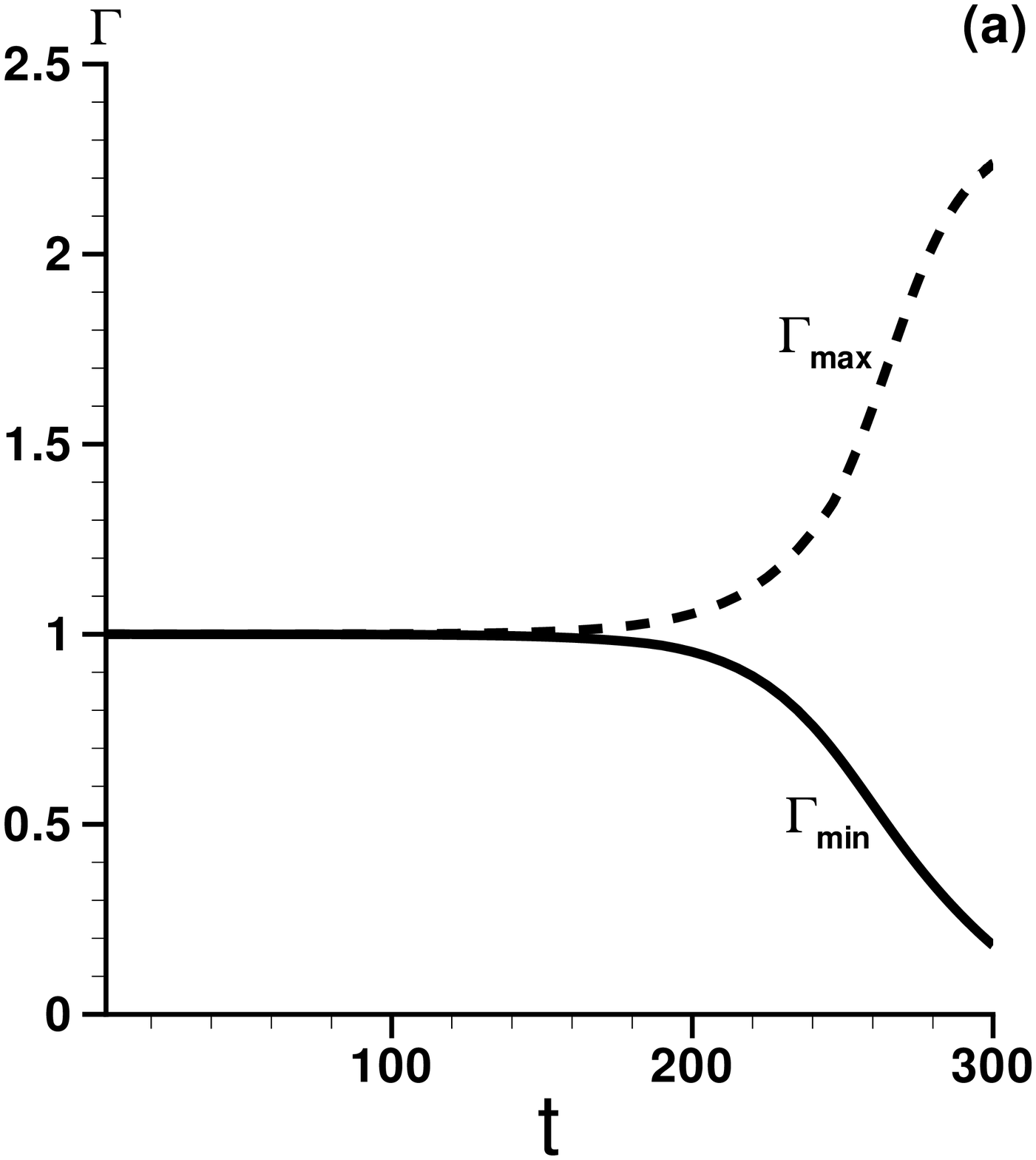}\includegraphics[%
  width=2.5in,
  keepaspectratio]{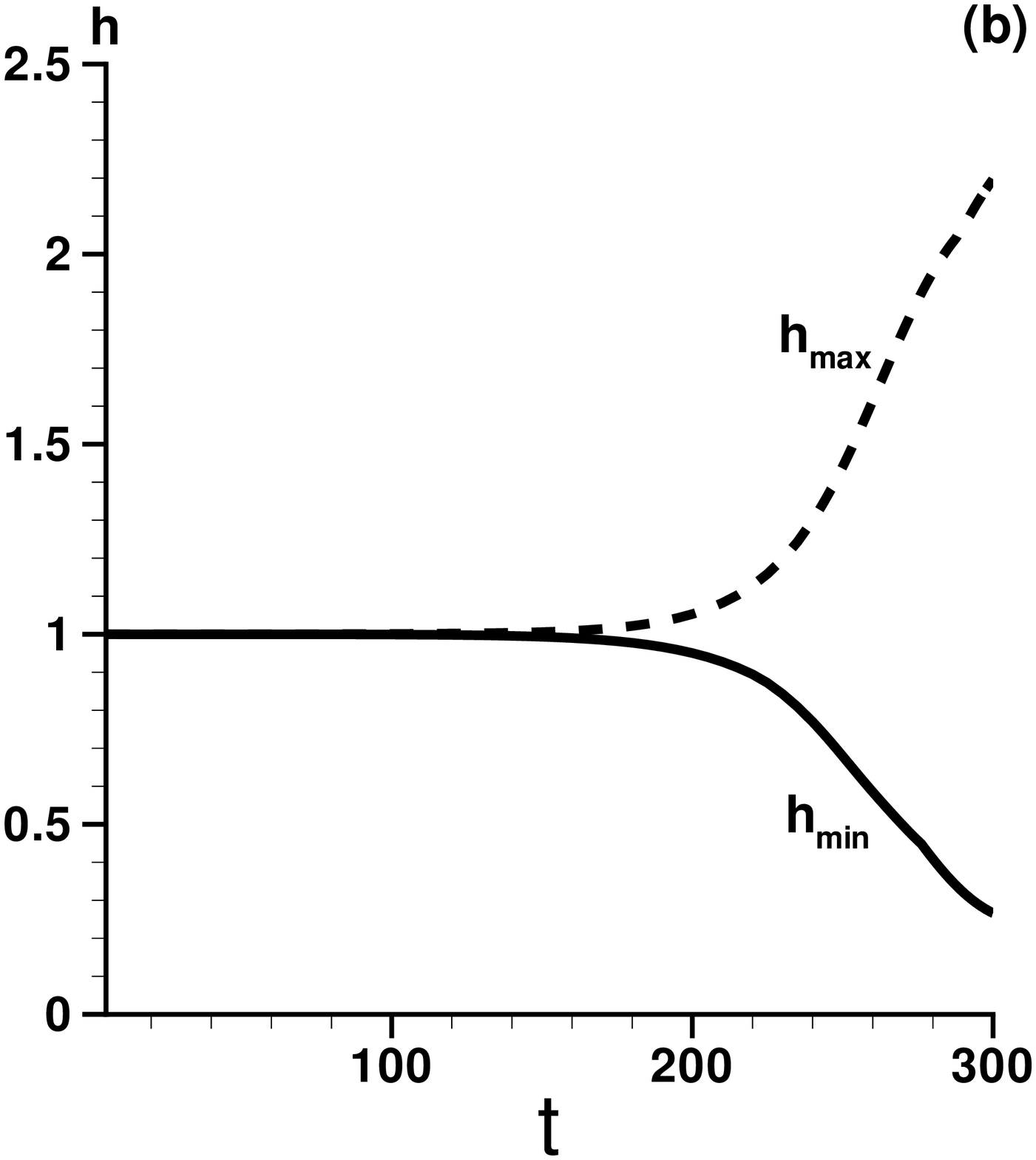}\end{center}

\caption{Same as Fig. 6 but for a non-zero diffusivity $D$ value. Here $D=0.1$,
$C=1$ ($Q=18$).}

\label{fig:evolminmax-c1dp1}
\end{figure}

The above results have been obtained discarding surfactant diffusion.
To check the continuity in the surfactant diffusivity at its zero
value, we should add into the left-hand side of equation (\ref{eq:evolg})
the term $-D\Gamma_{xx}$, where $D$ is a constant proportional to
$D_{s}$, the surface molecular diffusivity of surfactant. We have
simulated the modified equations with the values of $D$ as large
as $0.1$. The results turned out to be qualitatively the same as
with $D=0$ (see Fig.~\ref{fig:evolminmax-c1dp1}). 

\begin{center}
\textbf{5. Discussion and conclusions}
\end{center}

We have investigated the nonlinear stages of the insoluble-surfactant
instability in a sheared viscous film in a semi-infinite plane-Couette
flow. Numerical simulations of strongly nonlinear longwave evolution
equations on extended spatial intervals show that the earlier uncovered
remarkable property of the linear instability, the necessity of a
non-zero interfacial shear rate for the system to be unstable to infinitesimal
disturbances, has a {}``nonlinear continuation'': no matter how
large or small the interfacial shear of the base flow, there is \textit{no}
small-amplitude saturation of the instability. (So the weakly non-linear
equations cannot describe the large-time evolution.)

The impossibility of a small-amplitude saturation is corroborated
by the following argument. Suppose there is such a saturated state
of Eqs.~(\ref{eq:res-w-n-h}) and (\ref{eq:res-w-n-g}) with $N<<1$
and $G<<1$, where $N$ and $G$ are the saturated amplitudes of deviations
of the film thickness and surfactant concentration, respectively,
having the characteristic length scale $L$ (such that $\eta_{z}\sim N/L$
and $g_{z}\sim G/L$) and time scale $T$ (such that $\eta_{t}\sim N/T$
and $g_{t}\sim G/T$). Consider the dominant balance of terms in Eqs.~(\ref{eq:res-w-n-h})
and (\ref{eq:res-w-n-g}). The only nonlinear term, $\eta\eta_{z}$,
must be one of the dominant terms in Eq.~(\ref{eq:res-w-n-h}). But
at least one of the terms $Cg_{zz}$ and $C\eta_{zzzz}$ must be among
the dominant terms there, since the dominant equation cannot be just
$\eta_{t}+\eta\eta_{z}=0$: this equation is well-known to lead to
infinite slopes in finite time. Thus, $N^{2}/L\sim\text{max}(CG/L^{2},CN/L^{4})$.
But the same terms, $Cg_{zz}$ and $C\eta_{zzzz}$, cannot be among
the dominant ones in Eq.~(\ref{eq:res-w-n-g}), since the term $\eta_{z}$
there is much larger: $N/L>>N^{2}/L$. Thus the dominant balance in
Eq.~(\ref{eq:res-w-n-g}) would have to be just $g_{t}+\eta_{z}=0$,
leading to $\eta_{z}\sim g_{t}$. Combining this relation with $\eta_{t}\lesssim\eta\eta_{z}$
(see Eq.~(\ref{eq:res-w-n-h})), we can write $\eta_{t}\lesssim\eta g_{t}\sim NG/T<<N/T\sim\eta_{t}$.
Thus, we have arrived at $\eta_{t}<<\eta_{t}$, a contradiction. Therefore,
there can be no saturated solutions of the weakly-nonlinear equations
(\ref{eq:res-w-n-h})-(\ref{eq:res-w-n-g}) with both $\eta$ and
$g$ being small; in other words, there can be no small-amplitude
saturation of the instability in question. (This argument assumes a
single time scale, which is generically the case. Only for a limited
domain of near-critical states multiple time scales occur due to the
presence of the small super-criticality parameter.  As was mentioned
above, under such non generic conditions the small-amplitude
saturation does occur. This should be possible be describable by
Landau type theory, which we plan to investigate in the future.)

The same considerations hold for the weakly nonlinear equations obtainable
directly from Eqs.~(\ref{eq:hevol}-\ref{eq:gammaevol}). They imply
that, even when the Marangoni number $M$ is not small, the small-amplitude
saturation of the interfacial-surfactant instability is impossible
as well. 

Our numerics show that when one includes into consideration (small)
non-zero diffusion of the surfactant, the instability still does not
saturate with small amplitudes, and the disturbances of the film thickness
or/and the surfactant concentration reach values of order one. Evidently,
the increase of the base shear does \textit{not} lead to the small-amplitude
saturation--as it does in all previously known similar systems without
surfactants (which are described by a single evolution equation, the
one for the film thickness). Thus, for the interfacial surfactant
instability under consideration here, the increase of the saturating
effect with the shear rate cannot overcome the simultaneous increase
in the strength of the instability.

It remains to be seen if there is a threshold in the strength of the
surfactant diffusivity above which a small-amplitude saturation of
the instability will set in. We hope to fully investigate the effect
of diffusion (including the questions of saturation, with small as
well as with large amplitudes) in the future. Even without diffusion,
the large-amplitude saturation is possible; an example is shown in
Fig~\ref{fig:saturatedsol-cp01q5p2}. The systematic investigation
of the large-amplitude evolution, which is beyond the scope of this
paper, will be a subject of the future study as well.

\end{document}